%% file: paper.tex
\documentclass{sig-alternate-05-2015}

\input{preamble}

\newcommand{\sysname}{Ettu}
\newcommand{\featuresof}[1]{F_{#1}}
\newcommand{\digest}[1]{\texttt{digest}({#1})}
\newcommand{\listdigest}[1]{\texttt{digest}_{\ell}([{#1}])}
\newcommand{\bagdigest}[1]{\texttt{digest}_{b}(\{|{#1}|\})}
\newcommand{\setdigest}[1]{\texttt{digest}_{s}(\{{#1}\})}
\newcommand{\astof}[1]{\mathit{AST}_{#1}}

\setcopyright{acmcopyright}

\title{Summarizing Large Query Logs in \sysname{}\titlenote{The first three authors contributed equally and should be considered a joint first author}}

\numberofauthors{1} 
\author{
\alignauthor
Gokhan Kul, Duc Luong, Ting Xie, Patrick Coonan, \\
Varun Chandola, Oliver Kennedy, Shambhu Upadhyaya\\
       \affaddr{University at Buffalo, SUNY}\\
       \email{\{gokhanku, ducthanh, tingxie, pcoonan, chandola, okennedy, shambhu\}@buffalo.edu}
}

\toappear{}

\begin{document}

\maketitle

\input{sections/0-abstract.tex}

\keywords{Query Logs; Summarization; Database Security; Clustering}

\section{Introduction}
\input{sections/1-intro.tex}

\section{Background}
\label{sec:background}
\input{sections/background.tex}

\section{System Outline}
\label{sec:system}
\input{sections/2-system.tex}

\section{Clustering}
\label{sec:clustering}
\input{./sections/clustering}

\section{Summarization}
\label{sec:summarization}
\input{./sections/summarization}

\section{Experiments}
\label{sec:experiments}
\input{sections/3-experiments.tex}

\section{Related Work}
\label{sec:relatedwork}

\input{sections/4-relatedwork.tex}

\section{Conclusion and Future Work}
\label{sec:futurework}
\input{sections/5-futurework.tex}

\section{Acknowledgments}
This material is based in part upon work supported by the National Science Foundation under award number CNS - 1409551. Usual disclaimers apply. We also would like to thank Manish Gupta and Rick Jesse for their efforts in preparation of the dataset we used in our experiments.

{
\bibliographystyle{unsrt}
\bibliography{paper}
}

\end{document}

%% file: preamble.tex
\usepackage{graphicx}
\usepackage{amssymb}
\usepackage{algorithm}
\usepackage{algpseudocode}
\usepackage{textcomp}
\usepackage{listings}
\usepackage[usenames,dvipsnames]{xcolor}
\usepackage{subcaption}
\usepackage{cite}
\usepackage{hyperref}
\usepackage{lipsum}
\usepackage[normalem]{ulem}
\usepackage{listings} 

\newtheorem{example}{Example}
\newtheorem{scenario}{Scenario}
\newtheorem{definition}{Definition}

\newcommand{\comprehension}[2]{\left\{\;{#1}\;|\;{#2}\;\right\}}
\newcommand{\tuple}[1]{\left<\;{#1}\;\right>}



%% file: sections/0-abstract.tex
\begin{abstract}
Database access logs are large, unwieldy, and hard for humans to inspect and summarize.  In spite of this, they remain the canonical go-to resource for tasks ranging from performance tuning to security auditing.  In this paper, we address the challenge of compactly encoding large sequences of SQL queries for presentation to a human user.  Our approach is based on the Weisfeiler-Lehman (WL) approximate graph isomorphism algorithm, which identifies salient features of a graph --- or in our case of an abstract syntax tree.  Our generalization of WL allows us to define a distance metric for SQL queries, which in turn permits automated clustering of queries. We also present two techniques for visualizing query clusters, and an algorithm that allows these visualizations to be constructed at interactive speeds. Finally, we evaluate our algorithms in the context of a motivating example: insider threat detection at a large US bank. We show experimentally on real world query logs that (a) our distance metric captures a meaningful notion of similarity, and (b) the log summarization process is scalable and performant.
\end{abstract}

%% file: sections/1-intro.tex

Database access logs are used in a wide variety of settings, from performance monitoring, to benchmark development, and even to database auditing and compliance.  Examining a history of the queries serviced by a database can help database administrators with tuning, or help security analysts to assess the possibility and/or extent of a security breach.  However, logs from enterprise database systems are far too large to examine manually.  As one example, our group was able to obtain a log of all query activity at a major US bank for over a period of 19 hours. The log includes nearly 17 million SQL queries and over 60 million stored procedure execution events.  
Even excluding stored procedures, it is unrealistic to expect any human to manually inspect all 17 million queries.

Let us consider an analyst (let's call her Jane) faced with the task of analyzing such a query log.  She might first attempt to identify some aggregate properties about the log.  For example, she might count how many times each table is accessed or the frequency with which different classes of join predicates occur.  Unfortunately, such fine-grained properties do not always provide a clear picture of how the data is being used, combined, and/or manipulated.  To get a complete picture of the intent of users and applications interacting with the database, Jane must look at entire queries.  So, she might turn to more coarse-grained properties, like the top-k most frequent queries.  Here too she would run into a problem.  Consider the following two queries: 
\begin{verbatim}
           SELECT * FROM R WHERE R.A = 1
           SELECT * FROM R WHERE R.A = 2
\end{verbatim}
They differ only in the value of a single constant: 1 or 2.  When counting queries, should she count these as the same query or two different queries?  If she chooses to count them together, are there perhaps other ``similar'' queries that she should also count together, like for example:
\begin{verbatim}
           SELECT * FROM R WHERE R.B = 1
\end{verbatim}
Of course the answer depends on the content of the log, the database schema, and numerous other details that may not be available to Jane immediately when she sits down to analyze a log.  As a result, this type of log analysis can quickly become a tedious, time-consuming process.

In this paper, we introduce a framework that automatically creates compact, easy-to-consume summaries of large query logs.  The first part of this framework is an algorithm for extracting features from the abstract syntax trees (ASTs) of SQL queries.  The resulting feature vectors help to define a similarity metric for queries, which in turn allows us to cluster queries into groups based on their structure.

To further aid users in their understanding of query logs, we propose two techniques for summarizing and visualizing queries in a cluster: (1)  a text explanation that overviews common features of queries in the cluster, (2) a graph visualization that presents these features in the context of an actual query.

\subsection{Motivating Application: \sysname{}}
\label{sec:motivation}
It is increasingly important for organizations to be able to detect and respond to cyber attacks.
An especially difficult class of cyber attack to detect is the so called \textit{insider attacks} that occur when employees misuse legitimate access to a resource like a database.
The difficulty arises because apparently anomalous behavior from a legitimate actor might still have legitimate intent.  
For example, a bank teller in Buffalo who withdraws a large sum for a client from California may be acting legitimately (e.g., if the client has just moved and is purchasing a house), or may be committing fraud.
The ``U.S. State of Cybercrime Survey''~\cite{cybercrimeReport2014} states that 37\% of organizations have experienced an insider incident, while a 2015 study~\cite{ponemonReport2015} identified insider attacks as having the longest average response time of any attack type surveyed: 54.5 days.

The challenge of addressing of insider attacks lies in the difficulty of precisely specifying access policies for shared resources such as databases. 
Coarse, permissive access policies provide opportunities for exploitation. 
Conversely, restrictive fine-grained policies are expensive to create and limit a legitimate actor's ability to adapt to new or unexpected tasks.
In practice, enterprise database system administrators regularly eschew fine-grained database-level access control.  
Instead, large companies commonly rely on reactive strategies that monitor external factors like network activity patterns and shared file transfers.
In a corporate environment, monitoring user actions requires less preparation and gives users a greater degree of flexibility. 
However, external factors do not always provide a strong attestation of the legitimacy of a database user's actions. 

The \sysname{}\footnote{\sysname{} is derived from the last words of the Roman emperor Julius Caesar, ``\textit{Et tu, Brute?}'' in Latin, meaning ``\textit{You, too, Brutus?}'' in English to emphasize that this system is meant to detect the unexpected betrayals of trusted people} system, currently under development at the University at Buffalo~\cite{kul2016ettu}, seeks to help analysts to monitor query access patterns for signs of insider attack.  Database logging and monitoring is expensive, so \sysname{} needs to be able to identify normal, baseline database behaviors that can be easily flagged as ``safe'' and ignored from normal logging and post-mortem attack analysis.  In this paper, we focus on one concrete part of the overall \sysname{} system, responsible for summarizing and visualizing query logs.  In the complete system, this component serves to help analysts generate patterns of safe queries, and to quickly analyze large multi-day query logs to identify potential attack activity.

\subsection{Contributions}

In addition to enabling insider threat analysis~\cite{kul2016ettu}, the ability to quickly visualize patterns in query logs has potential applications for new benchmark development~\cite{pocketdata}, database tuning~\cite{Bruno:2005:APD:1066157.1066184}, and privacy analysis~\cite{Dwork2006}.  This paper represents a first step towards effective summarization for query logs.
Concretely, in this paper we:
(1) identify and motivate the challenge of log summarization,
(2) adapt the Weisfeiler-Lehman approximate graph isomorphism algorithm~\cite{WL2011} to the task of feature extraction from SQL queries,
(3) propose Euclidean distance over the resulting feature vectors as a similarity metric for SQL that can be used create clusters of structurally similar queries,
(4) introduce techniques for visualizing and summarizing the queries in each cluster, and
(5) experimentally demonstrate that our similarity metric mirrors human intuition and that our clustering and visualization processes are scalable.

This paper is organized as follows.  
We provide background information that creates a basis for our research in Section~\ref{sec:background} and introduce our core contribution, a technique for query similarity evaluation, in Section~\ref{sec:system}.
We next explain our proposed clustering and summarization techniques in Sections~\ref{sec:clustering} and~\ref{sec:summarization}, respectively. 
In Section~\ref{sec:experiments}, we evaluate the accuracy and performance of our proposed techniques. We compare our approach to related work in Section~\ref{sec:relatedwork}.
Finally, we conclude by identifying the steps needed to deploy query summarization into practice in Section~\ref{sec:futurework}.


%% file: sections/background.tex


As a declarative language, the abstract syntax tree (AST) of a SQL statement acts as a proxy for the intent of the query author.  
We thus argue that structural similarity is a meaningful metric for query similarity.
Naively, we would group a query $Q$ with other queries that have nearly (or completely) the same AST as $Q$.
This structural definition of intent has seen substantial use already, particularly in the translation of natural language queries into SQL~\cite{li2015NLPI}.  

\begin{figure}[h!]
\centering
\includegraphics[width=6cm]{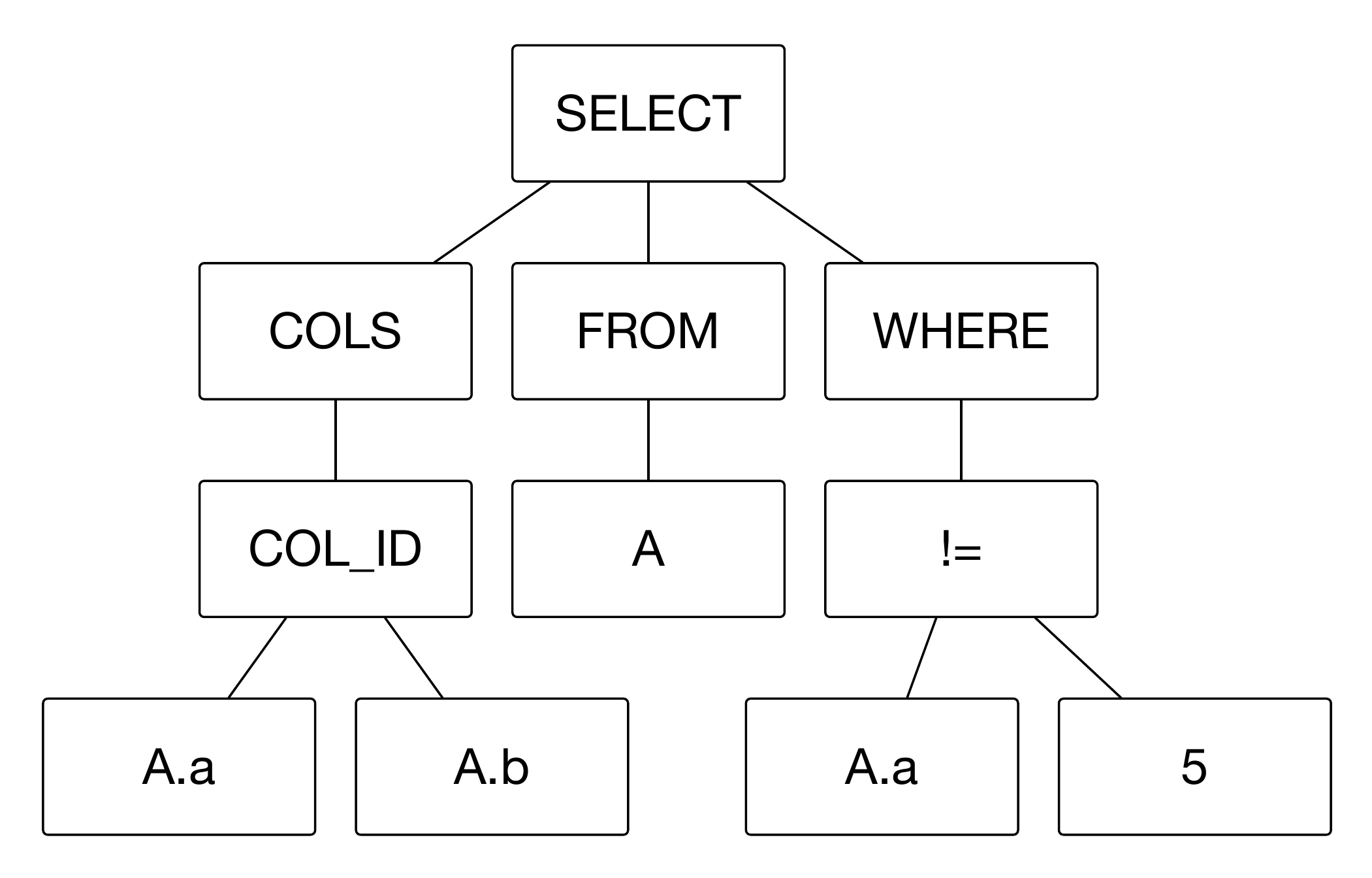}
\caption{An abstract syntax tree of query \texttt{SELECT A.a, A.b FROM A WHERE A.a != 5}.}
\label{fig:exampleAST}
\end{figure}

For the remainder of this paper, we will use $Q$ to denote both the query itself as well as its tree encoding.
An ideal distance metric would measure the level of similarity or overlap between these tree encodings and their substructures.  
Naively, for two SQL queries $Q_1$ and $Q_2$, one satisfactory metric might be to count the number of connected subgraphs of $Q_1$ that are isomorphic to a subgraph of $Q_2$.  
Subgraph isomorphism is NP-complete, but a computationally tractable simplification of this metric can be found in the Weisfeiler-Lehman (WL) Algorithm~\cite{WL2011} illustrated as Algorithm~\ref{alg:WLAlgorithm}.
Instead of comparing all possible subgraphs of $Q_1$ against all possible subgraphs of $Q_2$, the WL algorithm restricts itself to specific types of subgraphs.

\begin{algorithm}
\caption{Weisfeiler--Lehman Algorithm}
\label{alg:WLAlgorithm}
\begin{algorithmic}[1]
\Procedure{Weisfeiler\textendash Lehman}{}
\For{each tree $t \in T$ }
\State $H \gets depth(Q)$
\For{$h \gets 1; h \le H; h++$ }
\For{each node $N \in Q$ }
\State $s \gets CreateSet(labels of \{N\}~\bigcup \sum N.children)$
\If{$hashtable.get(s) \neq null$}
\State $label(N) \gets hashtable.get(s)$
\Else
\State $hashtable.put(s, label(N))$
\EndIf
\If {$IsLeaf(\sum N.children)$}
\State $N.IsLeaf = true$
\EndIf
\EndFor
\EndFor

\EndFor
\EndProcedure
\end{algorithmic}
\end{algorithm}


Given a query $Q$, let $N \in Q$ denote a node in $Q$.  $N$ is initially labeled with the SQL grammar symbol that $N$ represents.
The \textit{i-descendent} tree of $N$: $desc(N, i)$ is the sub-tree rooted at $N$, including all descendants of $N$ in $Q$ up to and including a depth of $i$.  

\begin{example}
Given the tree in Figure~\ref{fig:exampleAST}, 
$desc(\texttt{COLS}, 2)$ is the tree containing the nodes \texttt{COLS}, \texttt{COL\_ID}, \texttt{A.a}, and \texttt{A.b}.  
\end{example}
The WL algorithm identifies a query $Q$ by all possible i-descendent trees that can be generated from $Q$:
$$id(Q) = \comprehension{desc(N, i)}{N \in Q \wedge i \in [0, depth(Q)]}$$
Here $depth(Q)$ is the maximum distance from the root of $Q$ to a leaf. To make this comparison efficient, subtrees are deterministically assigned a unique integer identifier, and the query is described by the bag of $Q$'s i-descendent tree identifiers.  Thus two query trees with an isomorphic subtree will both include the same identifier in their description.
The bag of identifiers is encoded as a (sparse) feature vector and allows Euclidean distance to measure the similarity (or rather dis-similarity) of two queries.

%



\begin{figure}[h!]
\centering
\includegraphics[width=\columnwidth]{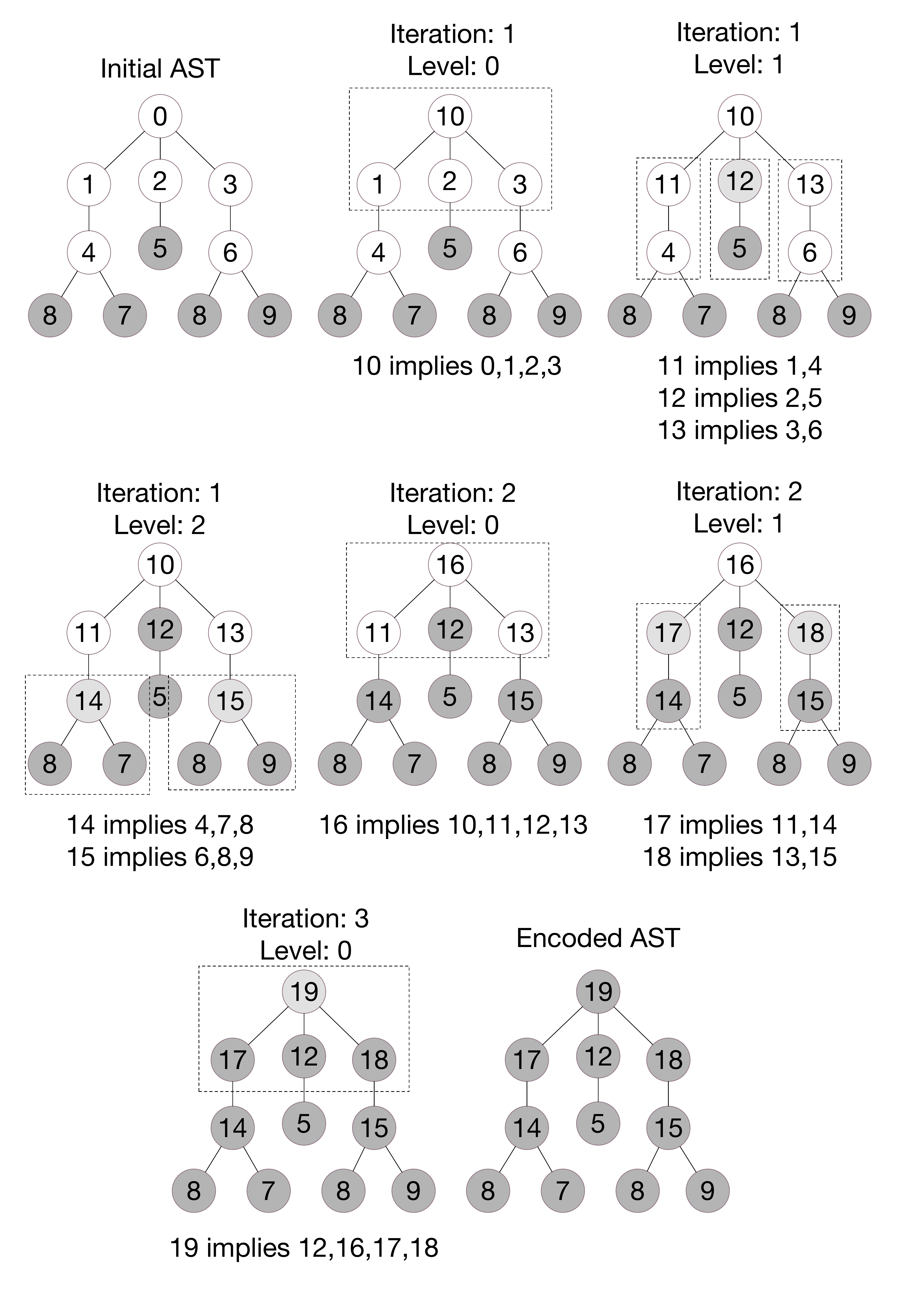}
\caption{Weisfeiler-Lehman algorithm applied on AST given in Figure~\ref{fig:exampleAST}.}
\label{fig:wlexample}
\end{figure}

Figure~\ref{fig:wlexample} shows how the WL algorithm is applied on the AST of the query given in Figure~\ref{fig:exampleAST}. First, every distinct node of the AST gets labeled with a unique integer.  If the same node appears more than once, each instance is labeled with the same integer.
As the algorithm progresses, the dotted box emphasizes the region being examined, while the text below represents new labels being synthesized from existing labels.  Grey nodes have been fully labeled, while white nodes are still being processed.

%% file: sections/2-system.tex

\begin{figure}[h!]
\centering
\includegraphics[width=\columnwidth]{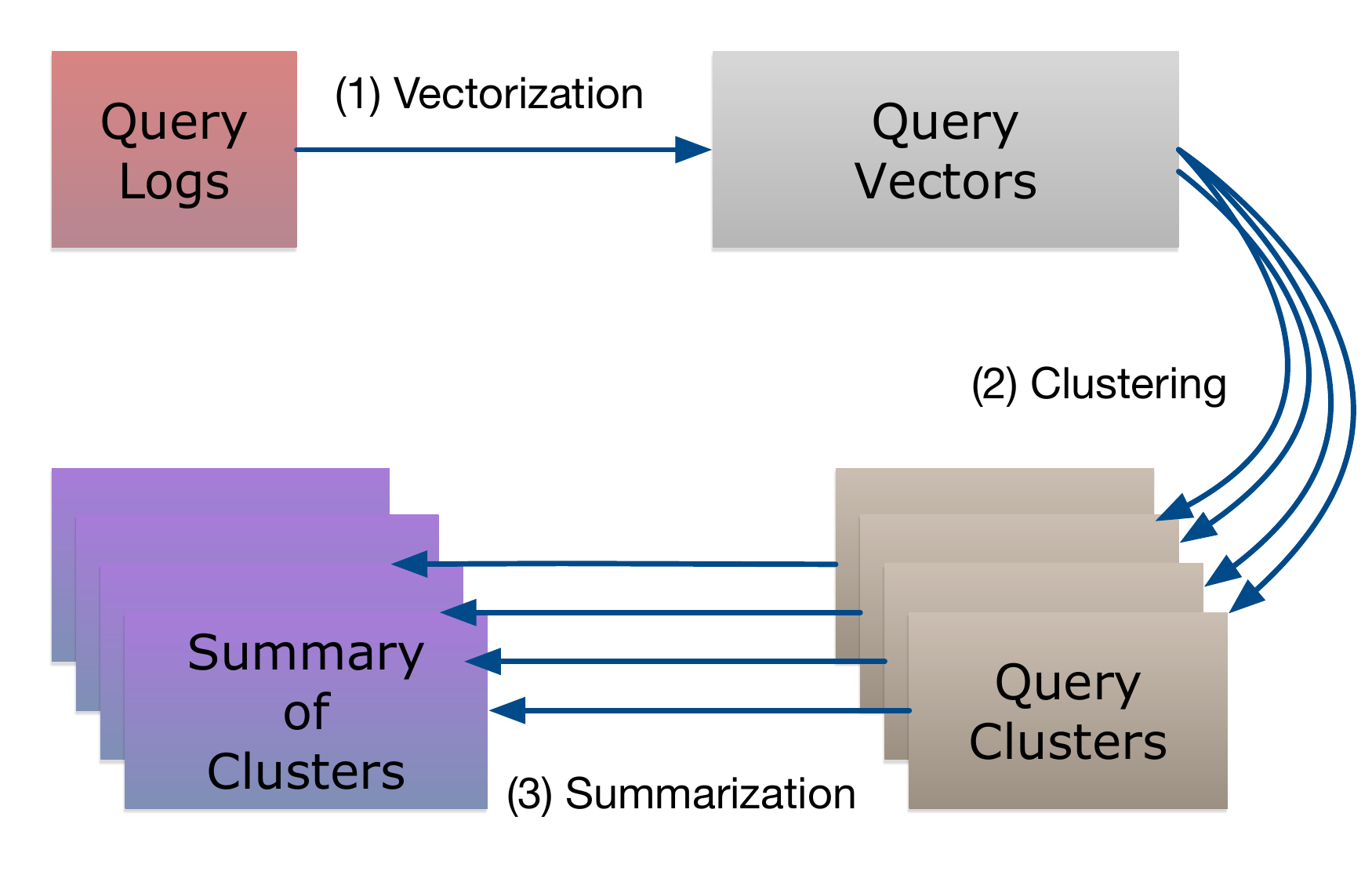}
\caption{The typical query auditing workflow in \sysname{}.}
\label{fig:workflow}
\end{figure}

The log summarization mechanism in \sysname{} operates in three stages: (1) A \textbf{vectorization} phase where the aggregated query logs are processed into feature vectors, (2) An offline \textbf{clustering} phase where the feature vectors are grouped together according to their structural similarity, and (3) A \textbf{summarization} phase where the clusters are summarized to present a human user with a concise overview of the cluster's distinguishing features.  These stages are illustrated in Figure~\ref{fig:workflow}.

The initial input to \sysname{} is a log of query activity processed by the target database. The log may be annotated with supplemental metadata like usernames and timestamps.
The goal of the methodology described here is to produce a concise, but precise sketch of the most common query structures in the log.
This summary contains information about groups of similar queries aiming to explain what a query in each group looks like and what kind of features are expected to be encountered in a query.

\begin{scenario}
Jane feeds the logs collected to \sysname{}. The system processes the file and groups similar queries together and passes them to the next stage of the operation.
\end{scenario}

\subsection{Generalization of Weisfeiler-Lehman}
\label{GenOfWL}
The WL algorithm assumes zero knowledge about structural features of the trees it compares, limiting itself to i-descendent subtrees.  
Conversely, the grammar of SQL has a very well-defined structure.  
In \sysname{}, we exploit this structure to eliminate redundancy and create features that more reliably encode the query's semantics.  
We specifically identify and address three features of SQL that limit the effectiveness of WL.

First, the number of features created by the WL algorithm is large. Although clustering naturally prunes out features without discriminative power, we can use SQL's semantics to identify structures that are unlikely to be useful.  For example, the subtree $desc(\texttt{SELECT}, 1)$ in Figure~\ref{fig:exampleAST} is common to virtually all queries. Such features can be pruned preemptively.

Second, constants in the query require special treatment in feature encoding. The output of this operation is a \textit{query skeleton}.
\begin{example}
\label{example: skeleton}
Consider the query \texttt{SELECT * FROM R WHERE R.a = 1}.  Subtree \texttt{R.a = 1}  can be abstracted as \texttt{R.a = ?} because there is much higher probability that this abstraction will be shared thus it correctly captures commonality among queries while feature \texttt{R.a = 1} can coexist offering distinguishing power ( \texttt{?} denotes the placeholder constant)
\end{example}

Finally, SQL makes frequent use of commutative and associative operators.  
The semantics of such operators may overlap, even if their i-descendent subtrees do not. 
\begin{example}
Consider the Boolean expressions \texttt{A AND B} and \texttt{A AND B AND C}.  The former AST is a \texttt{AND} node with 2 children while the latter has 3.  Although the two ASTs have 3 0-descendent subtrees in common, they share no 1-descendent subtrees; WL ignores the similarity between the two conjunctive expressions.
\end{example}
To address these challenges we first generalize WL algorithm to allow the creation of dynamic, language-specific features, and then use this facility to present a suite of features that permit the generalized W-L algorithm to reliably detect features of SQL queries.

\subsection{Dynamic Features}
\label{sec:features}
Feature generation process can be explained in three steps:
(1) Based on the structure of the query AST and existing set of features, use some \textit{rules} to identify collections of existing features as candidates for new feature creation;
(2) Use a set of \textit{constraints} to filter out undesirable features;
(3) Offer a function $\digest{\cdot}$ to synthesize new feature labels from these collections of existing features.
To be more precise, suppose we are given a query $Q$, each node $N \in Q$ has a set of \textit{features} $\featuresof{N}$, initially defined as the singleton set containing its SQL grammar atom.
A \textit{rule} is applied to construct $\featuresof{N}$ from the node itself, features generated by other rules, and the feature sets of its descendants: $\comprehension{\featuresof{C}}{\text{C is a descendent of N}}$.
A \textit{constraint} acts as a filter, removing features from consideration.
Rules and constraints are applied in a pre-defined priority order, bottom-up to the nodes of $Q$ to compute feature sets for each of $Q$'s nodes.  
Finally, the feature vector of $Q$ is defined by the bag $\biguplus_{N \in Q}\featuresof{N}$ of all features from all nodes of $Q$.
Note that each node has at most one instance of a feature ($\featuresof{N}$ is a set), while an entire AST $Q$ might have multiple instances of the feature (the feature vector of $Q$ is a bag).

Let us consider the steps of feature generation in reverse order.  First, to define features by structural recursion, we need a way to synthesize new feature identifiers from collections of existing identifiers.  We define a function $\digest{\cdot}$ that deterministically constructs feature identifiers from existing collections (whether bags, sets, or lists) of identifiers.  The output of $\digest{\cdot}$ is deterministic and unique: for any collection $A$, $\digest{A} = \digest{A}$ and for any two collections $A \not \equiv B$, $\digest{A} \neq \digest{B}$.  Uniqueness is achieved by maintaining a persistent table of all identifiers assigned, while determinism for bags and sets is achieved by first sorting features in the collection.  We will denote list, bag, and set digests as $\listdigest{a, b, a, c}$, $\bagdigest{a, a, b, c}$, and $\setdigest{a, b, c}$, respectively.  

\begin{example}
List-digests respect order, and as a result $\listdigest{a,b,a,c} \neq \listdigest{a,a,b,c}$.  Bag-digests respect multiplicity but not order, so $\bagdigest{a,b,a,c} = \bagdigest{a,a,b,c}$, but also that $\bagdigest{a,a,b,c}\neq \bagdigest{a,b,c}$.  Finally, set digests differentiate neither, so $\setdigest{a,a,b,c} = \setdigest{a,b,c}$
\end{example}

Second, we wish to ignore features that are semantically uninformative, like the 1-dependent subtree $desc(\texttt{SELECT}, 1)$ in Figure~\ref{fig:exampleAST}.  Given a list of these features and/or patterns describing these features, we can mark unwanted features by in the table of identifier assignments generated by $\digest{\cdot}$.  Unwanted feature identifiers are simply not added to $\featuresof{N}$ as it is constructed.


Finally, we consider the construction of new features by structural recursion.  We have found that all of the customized features we are interested in creating can be expressed through first--order logic, or more precisely though datalog~\cite{huang2011datalog}. In datalog, logic rules are expressed in the form
$$Head(Y) :- Body(XY) \equiv \forall y\exists x \left(Body(xy) \rightarrow Head(y)\right)$$
Here, $Body(XY)$ is a conjunction of first--order predicates over act as conditions, true for some set of valuations to the variables $XY$.  The rule indicates that the conjunction of predicates $Head(Y)$ should be true for any valuation for which there exists some $X$ for which $Body(XY)$ is true.  

For rule generation, we employ two types of predicates: 
\begin{itemize}
\item Structural predicates, which are defined according the AST's graph structure. We use two structural predicates: $Node(N)$ is true for all nodes $N \in G$ and $Child(N_{c}, N_{p}, n)$ is true if $N_c$ is the $n$-th child of $N_p$.
\item Feature predicates,which are defined by the rule system. The feature predicate $Atom(N,f)$ is true if $f$ is the feature encoding the grammar atom for node $N$ in the AST. Rules define new features by populating predicate $Own(N, f)$ with $Atom(N,f)$ as initial elements $N$ would own.
\end{itemize}
For easy explanation, we present Table~\ref{rule_predicates} with built-in predicates that we will use in our feature generating rules. We will first show that Weisfeiler-Lehman algorithm can be easily re-formalized under datalog using recursion in~\ref{sec:NewWL}. Finally, in \ref{sec:skeleton}-\ref{sec:CNF}, the solutions for challenges brought by WL algorithm are re-formalized in datalog.

\begin{table}
\small
\centering
\begin{tabular}{| p{2.2 cm}  | p{5.2cm} |}
\hline 
Predicate               & Description                                                                 \\\hline
$Node(N)$             & Node $N$ in query AST.               \\\hline
$Atom(N,f)$         & Feature $f$ encodes the grammar atom of node $N$.                  \\\hline
$Child(N_c,N_p, n)$   &  Node $N_c$ is $n^{th}$ Child of $N_p$.  \\\hline 
$Own(N, f)$ & $N$ owns $f$ or equivalently Feature $f$ is in set $F_N$ of Node $N$.  \\\hline                    
\end{tabular}
\caption{Dynamic Feature Selection Predicates.}
\label{rule_predicates}
\end{table}

\subsubsection{Weisfeiler-Lehman Revisited}
\label{sec:NewWL}

We first show that the base Weisfeiler-Lehman algorithm can be easily re-formalized under datalog with recursion and aggregation.  In practice, the model of datalog that we use is loosely based on that of LogiQL~\cite{Aref:2015:DIL:2723372.2742796} which supports aggregate and grouping operations.

%
%
\smallskip
\noindent 1. $IterF(N, f_N, 0) :-\; Atom(N,f_N)$\\
2. \textbf{with} $f = \listdigest{ f_N, \bagdigest{ \forall f_C  } }$\\
\hspace*{0.33cm} $IterF(N,f ,n) :-\; $ \\
\hspace*{1cm} $IterF(C, f_C, n-1)$, $Child(C, N, \_)$, \\
\hspace*{1cm} $Atom(N,f_N)$, $(n<Height(N))$\\
3. $Own(N, f) :-\; IterF(N, f, \_)$
\smallskip

Predicate $IterF(N,f,n)$ is an intermediate store for features $f$ created for node $N$ at iteration $n$. $\_$ denotes a wildcard that accepts anything and we denote aggregate grouping by a $\forall$ symbol. Line 1 serves as initial condition. Starting from line 2, the datalog expression forms a group of all child labels $f_C$ assigned during the prior iteration and computes a bag-digest over the result ($\bagdigest{\forall f_C}$).

\subsubsection{Skeletons}
\label{sec:skeleton}
To give special treatment to binary expressions which we only care about its left hand side, Example~\ref{EqualitySkeleton} shows a rule that works on Equality operator.
\begin{example}
\label{EqualitySkeleton}
Name : Equality Skeleton

\textbf{with} $f=\listdigest{f_p, f_l, ?}$

$Own(f,N_p)$ $:-\;$

\hspace*{1cm} $Atom(N_p,f_{p})$, $N_p=Equals$,

\hspace*{1cm} $Child(N_r, N_p,2)$, $Atom(N_r, f_r)$, $Const(f_r)$,
 
\hspace*{1cm} $Child(N_l, N_p,1)$, $Atom(N_l, f_l)$, $\neg Const(f_l)$ 

\end{example}

$?$ is a placeholder indicating a constant value. $Const$ is a unitary predicate indicates whether a string represents a constant value. This datalog expression seeks any node $N_p$ in the AST whose grammar atom is $Equals$ and whose right child's atom is a constant. Right child is replaced by $?$ and $\listdigest{f_p, f_l, ?}$ generates the resulting feature that encodes the 1-descendent subtree AST rooted at $N_p$. 

\subsubsection{Conjunctive Normal Form}
\label{sec:CNF}

In response to the challenge brought by commutative and associative operators, one solution would be pre-processing the query AST and converting the composite boolean formula into \textit{Conjunctive Normal Form} (CNF). Since boolean formula in the query reveal the intent by which the query issuer defines and limits the desired data records, algebraically equivalent boolean formula reflect the same intent. Thus CNF normalization which re-formalizes an equivalence class of boolean formula into a uniform structure helps to generate features more informative in the sense of intent. CNF is preferred over Disjunctive Normal Form(DNF) as CNF is structurally more sensitive to logic OR expressions which is one of the major tools of retrieving additional data records. For example, the structure of OR-piggybacked formula ((A AND B) OR C) in its CNF changes to ((A OR C) AND (B OR C)) comparing with AND-piggybacked formula (A AND B AND C).
In this section, we show that pre-processing the subtree of some boolean formula into its CNF can be translated into rules expressed in datalog.

First, we introduce the basic predicates and notations used in this section. 
CNF is a conjunction of clauses in which each clause is a disjunction of \textit{literals}.
A literal is an atomic boolean formula or its negation. Accordingly, we define unitary predicate $L(N)$ to indicate that node $N$ is the root of the subtree in the AST that represents a literal. 
Any literal in a composite formula can be located by seeking a boolean operator node with zero logic connectives within its operand's subtree. Hence the predicate $L(N)$ can be realized by creating rules that enumerate all root nodes of literals in a boolean formula. As the AST subtree of a literal can be fully represented by its root, we give it the special notation $N_L$ that distinguishes it from common node notation. 

Logic OR operator is commutative, hence a Disjunctive Clause denoted by $DC$ in CNF is a set of literals. The predicate that records the member literals of a $DC$ is denoted as $Member(N_L,DC)$. For simplicity, we use notation $DC$ and $N_L$ representing both the object and the unique ID of the object. Identification of $DC$ is generated by digesting all member IDs in its set, denoted as $\setdigest{ \bigcup_{N_L \in DC} N_L}$. 

Since logic AND operator is also commutative, the CNF of any formula rooted at node $N$ can be viewed as a set of $DC$s in which each member is attached to $N$. We use a new predicate $Attach(DC,N)$ indicating that $DC$ is attached to $N$. This predicate enables \sysname{} to attach the generated CNF structure to the root of the original boolean formula.
If negated formula are considered, the rule for CNF normalization should be paired with the rule for DNF normalization. Turning a negated formula into CNF is the same as turning it into its DNF then its CNF is equivalent to a conjunction of negated conjunctive clauses of its DNF. 
For simplicity we assume there is no negated formula and all nodes in the example \ref{CNF} is a descendent of the root of some target boolean formula.

\begin{example}
\label{CNF}

Name: Conjunctive Normal Form

(1) Base case:   \textbf{with} $DC=$$\setdigest{N_L}$ 
 
$Member(N_L,DC)$, $Attach(DC,N_L)$  $:-\;$ $L(N_L)$

(2) For AND node:

$Attach(DC,N)$ $:-\;$ 

\hspace*{2.5cm}$Atom(N,f)$, $f=AND$, 

\hspace*{2.5cm}$Attach(DC,N_c)$, $Child(N_c,N,\_)$

(3) For OR node:

  \textbf{with} $DC=$ 
  
  $\setdigest{\bigcup_{ Member(N_L, DC_{m}) \lor Member(N_L ,DC_{n})}  N_L }$

$Member(N_p,DC)$,$Member(N_q,DC)$,

$Attach(DC,N)$ $:-\;$ 

\hspace*{0.4cm} $Member(N_p,DC_{m})$,$Attach(DC_{m},N_i)$,$Child(N_i,N,i)$, 

\hspace*{0.4cm} $Member(N_q,DC_{n})$,$Attach(DC_{n},N_j)$,$Child(N_j,N,j)$,

\hspace*{3.3cm} $Atom(N,f)$, $f=OR$, $(i\neq j)$

\end{example}

Since generating CNF for the sub-formula rooted at any node only requires information from its children, the CNF normalization process is bottom-up. For the base case, expression (1) in the rule creates a singleton bag containing a chosen literal itself as a new disjunctive clause and attach it to the root node of the literal. As the process is bottom-up, nodes assume that each of their children have already completed CNF and have the resulting disjunctive clauses attached to it. For root nodes of literals, since they do not have sub-formula, they only need to provide their CNF upwards while root nodes of logic connectives AND and OR need only combine children CNFs with respect to their own logic grammar. Expression (2) locates a logic AND node and attaches all disjunctive clauses from its children to itself. However, in expression (3) for logic OR node, we need to apply distributive law as a disjunction of CNFs need to be re-formalized as a single CNF. We union any pair of literal sets ($DC_{m}$, $DC_{n}$) attached to different children $N_i$,$N_j$ with $i \neq j$ of logic OR node $N$ and the resulting disjunctive clause $DC$ is attached to $N$. 







%% file: sections/clustering.tex
The next step is to cluster query skeletons by their feature vectors. The result of this process is a set of clusters in which query skeletons with similar structures are grouped together. We considered two possible clustering approaches for use in \sysname{}: k-means and hierarchical clustering~\cite{xu2005survey}.
K-means outputs a set of query skeletons for \textit{k} clusters. 
On the other hand, hierarchical clustering outputs a dendrogram -- a tree structure which shows how each query can be grouped together. 
For our specific use case where we aim to summarize groups of similar queries, the number of groups or clusters is not known in advance. For this reason, we eventually elected to use hierarchical clustering because it gives us the flexibility to choose the number of clusters after computing the dendrogram. 
In addition, a dendrogram is a convenient way to visualize the relationship between queries and how each query is grouped in the clustering process. 

Hierarchical clustering recursively selects the closest pair of data points and merges them into a single virtual data point.  This process requires an agglomerative rule, a formula for aggregating feature vectors together into the vector of each virtual data point.
There are several often used possibilities, including complete--linkage, single--linkage and average--linkage.
In our experiments, the choice of agglomerative rules does not significantly change the clustering result; all commonly used agglomerative rules produce equally reliable outputs.
Consequently, we arbitrarily chose to use complete--linkage, in which the farthest distance from a data point in one cluster to a data point in another cluster is considered as the distance between two clusters. 

Another aspect that we should consider when doing hierarchical clustering is is the distance metric to compare a pair of data points. 
In our experiments, we use Euclidean distance as the distance metric, though other metrics such as the Manhattan distance gave similar clustering outputs.


%% file: sections/summarization.tex


To help human inspectors reliably visualize and interpret clusters, we need a step to present sets of similar queries as compact summaries. An ideal presentation is in tree-like form: presenting a coarse view of the cluster at first, while allowing the inspector to easily unfold portions of the tree to obtain more detail.
As a basis for this visualization, we observe that each feature is constructed, essentially as a digest of grammar atoms for a syntactically correct \textit{partial} AST, and can be displayed as a subtree.

Each cluster can be characterized by a set of features common to most, if not all of the queries grouped into it.  An ideal form of summarization then, would first present a set of characteristic features for the entire cluster, and then allow users to iteratively subdivide the cluster into smaller and smaller sub-clusters, each with a progressively larger set of characteristic features.  We call the resulting hierarchical visualization a recursive feature characterization(RFC).

Generating a RFC requires us to overcome three challenges:  First, allowing a user to navigate through the RFC at an interactive speed requires scalable memory consumption and the ability to quickly identify clusters with large numbers of common features.  Second, the feature vectors used for clustering are usually redundant; The same node may appear multiple times across features representing distinct subtrees of the AST.  We need a way to quickly and efficiently identify and remove redundant features.
Overcoming the two above challenges is sufficient to produce a list of features that can be presented to a user. The first and most basic visualization is: a text-based summary of each cluster.  However, seeing features in isolation without the context of a surrounding query can make it difficult to understand the implications of each feature.  Consequently, our third challenge is to create a visualization that can present feature bags in context.

To address the first challenge, we adopt Frequent Pattern Trees (FP Trees)~\cite{han2004mining}.
Normally, FP Trees is used to mine frequent patterns of item-sets, for example sets of items frequently bought together.
Individual items in each item-set are sorted into a sequence and an FP Tree is built by prefix-aligning these sorted sequences. 
Any path of nodes starting from the root of the tree to an internal node in FP Tree represents a shared prefix of item sequences with each node counting the number of prefixes that pass it.
Each feature $f$ is constructed as the digest of a set of nodes of the query AST, the lineage of the feature.  The lineage can be reassembled into a partial AST, denoted by $\astof{f}$.  The multiplicity of $f$ in the feature bag $\biguplus_{N \in Q}\featuresof{N}$ generated for query $Q$ is denoted by $mult_Q(f)$ . For each query we create an item-set by constructing \textit{2-tuples} from features in the query $\comprehension{\tuple{f, mult_Q(f)}}{f \in \biguplus_{N \in Q}\featuresof{N}}$.  The resulting sets of items as 2-tuples are used to build an FP-Tree for the cluster.

\begin{figure}[h!]
\centering
\includegraphics[height=5 cm]{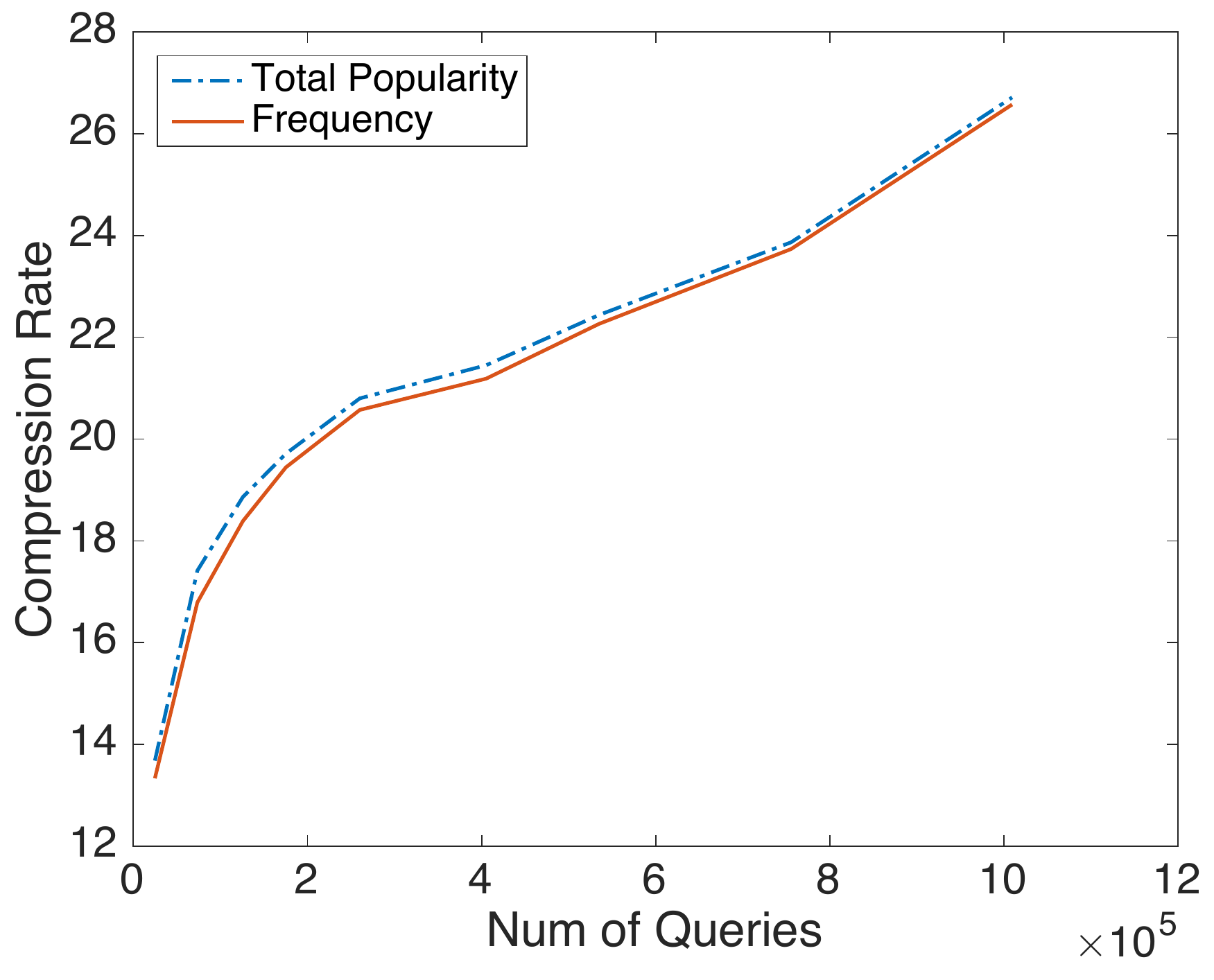}
\caption{Comparison of compression rate of Total Popularity and Frequency in FP tree.}
\label{fig:compression_rate}
\end{figure}

FP Trees offer high compression rates; the ratio of the size of an FP Tree to the total number of items appearing in the input item-sets is low.
To increase the compression rate, the item-sets can be sorted in the descending order of the number of occurrences of each feature tuple.  Doing so increases the chances that item-set prefixes can be shared, farther increasing the compression rate. 
This assumption has been validated experimentally~\cite{han2004mining} but can still fail, for example when an item occurs frequently, but as an isolated singleton.  As an optimization to counteract this case, we considered a metric related to frequency that we call \textit{total popularity}.  The \textit{popularity} of a feature in the feature vector of some query is the number of distinct features that coexist with it. \textit{Total popularity} is the sum of \textit{popularities} for the feature across all queries in the cluster. 
We compare the compression rate for FP Tree generated using both frequency and total popularity as sort orders in Figure~\ref{fig:compression_rate}.  
The graph shows how compression rate varies with the number of queries incorporated into the tree. 
As the number of queries increases, the compression rate for both approaches increases super-linearly, suggesting that FP Trees are a good way to visualize large query clusters, while the use of total popularity as a sort order provides a small, but measurable improvement in compression rate.

The FP Tree structure provides a tree-like sub-clustering of queries that naturally tracks features common to queries in the sub-clusters.  Users can selectively traverse it by folding or unfolding subtrees to settle on an appropriate level of detail, without being overwhelmed.
Recall that each path from the root to a current visited node in the FP tree corresponds to a prefix of item sequences shared by real life queries. 
As the user traverses the tree, \sysname{} tries to reduce the length of the prefix by assembling its underlying feature ASTs and ideally creates a single assembly AST with updated multiplicity as the summary. In practice, the resulting summary is a set of assembly ASTs and we would like to improve readability by reducing the set size. Further expanding and visualizing an FP Tree is as intuitive as simply merging more feature AST components to the existing assembly AST. 

\begin{figure*}[ht]
\centering
\includegraphics[height=11.5 cm]{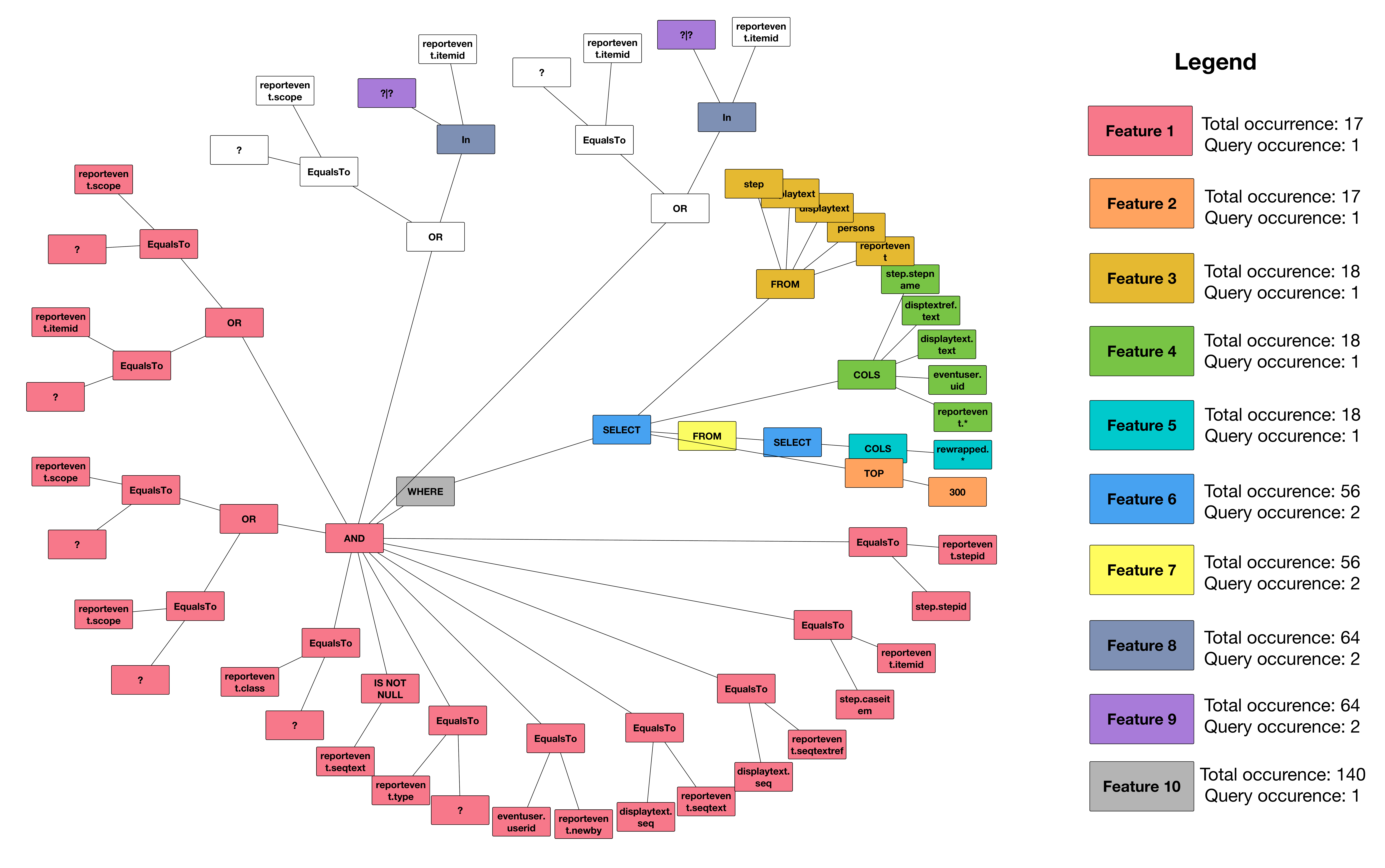}
\caption{An example AST of a SQL query where the common features that appears in all queries in the query group are colored and the uncolored nodes are specific only to that query.
}
\label{fig:summary_ast}
\end{figure*}

In general, reducing the set size contains two steps: (1) Shrink and (2) Merge.  The shrink step removes redundant items from the set as a preparation for merge step. Recall that an item $I$ is of the form $\tuple{f,mult_Q(f)}$. Given a specific feature $f_c \in \biguplus_{N \in Q}\featuresof{N}$, the set $S_c$ is defined as $$\comprehension{\tuple{f_p, mult_Q(f_p)}}{\astof{f_c} \subset \astof{f_p}, f_p \in \biguplus_{N \in Q}\featuresof{N}}$$ covers all items in the bag whose $\astof{f_p}$ fully contains $\astof{f_c}$. For any item $I_p=\tuple{f_p, mult_Q(f_p)} \in S_c$, suppose $f_c$ occurs $m_c$ times in the AST of feature $p$, by viewing item $I_p$, the inspector has already been informed of feature $f_c$ ($mult_Q(f_p)*m_c$) number of times. If we sum this for all $I_p \in S_c$ and compare the sum with the multiplicity of feature $f_c$, namely if $mult_Q(f_c) < \sum_{I_p\in S_c} mult_Q(f_p)*m_c$, item $I_c$ can be safely removed as redundant. Example \ref{shrink} shows a simple case of shrink.
\begin{example}
\label{shrink}
Consider a feature $f$ that encodes the expression $A=B$ with multiplicity 2, $\tuple{f,2}$. Consider another 2-tuple $\tuple{f',1}$ in the same feature bag which encodes $A=$ with multiplicity 1. It is possible that the multiplicity of $A=$ is less than $A=B$ as it is not necessary that each subgraph of $A=B$ is included in the bag. By viewing $A=B$ two times, the inspector is already informed of the existence of $A=$ structure two times which renders $\tuple{f',1}$ redundant.
\end{example}

Merge step further reduces the set size by a sequence of pairwise merge operation on items. Merge of two items is not merely the merge of their feature ASTs. The order of merge operations and the difference in feature multiplicity between two items should be carefully taken care of. In fact, the correctness of merge step needs to be verified by counting the number of original queries that share the resulting set of ASTs. This count should match with the count of queries, recorded in the FP Tree, that pass the chosen prefix.

In practice, shrink step is much faster and reliable than merge step: to guarantee the correctness of merge step, verification back on the original queries is required while shrink only involves local comparison on subtrees. In some cases that shrink step already produces minimal set size, merge step can be skipped.

Thus we recommend a property called \textit{Seal} for feature bags. Part of its goal is skipping merge step.
\begin{definition}
Property: Seal

Generic Seal: At least the complete subtree rooted at each node in the AST should be encoded as a feature; 

Tailored Seal: For any pair of distinct features in the feature bag of $Q$, if they assemble to $\astof{asmbly}$ which is also contained in query AST of $Q$, then structure $\astof{asmbly}$ should be included in the feature bag for $Q$.
\end{definition}
Generic Seal guarantees that features generated fully reconstructs the original query AST. 
With Tailored Seal, merge step can be skipped and features that encode assembly trees described in Tailored Seal are kept for summarization. Thus we are capable of providing the inspector a quick and a concise summary. It comes with the cost that depending on the number of additional features created, Seal would affect the scalability of \sysname{} with respect to memory consumption of FP Tree and speed of clustering. It may also affect clustering result as it changes the feature vector. Measuring and understanding the overall effect of applying property Seal are left for future work.

Figure~\ref{fig:summary_ast} shows an example AST resulting from a rule that only encodes the parental relationship between any node in query AST and the whole subtree rooted at one of its child node.
The AST is generated from a query and summarization information of the query set that it belongs to.
The 10 features shown with colored nodes appears at least once in every query in the query set. The legend shows that how many times the feature appears in all queries as ``Total occurrence", and in the example query itself as ``Query occurrence."
What makes this query different from the other ones is the uncolored nodes.
They either don't appear in the other queries, or appear infrequently.
This visualization makes a human inspector's work easier to spot important features and understand what the queries in the groups do, and identify why a query is different from the other queries.

%% file: sections/3-experiments.tex
In this section, we run experiments on our proposed framework to verify its performance and feasibility.
We are specifically interested in two properties of \sysname{}'s log classifier.  
First is whether or not our system can capture the notion of query intents through clustering queries by their similarity levels.
To answer this question, we ran an experiment to measure the level of correspondence between clustering results from \sysname{} and one from manual human inspection. 
As we will show, there is a high correlation between \sysname{} and manual clustering.
Second, especially given that clustering in general has $O(N^2)$ scaling in the data size, we want to show that \sysname{}'s clustering performs well enough to be practical for use in corporate settings.
As shown in the experiments, \sysname{} can cope very well with large data sizes even on commodity desktop computers.


The data we use to run our experiments is based on SQL query logs that capture all query activity on the majority of databases at a major US bank over a period of approximately 19 hours.  Logs are anonymized by replacing all constants with hash values generated by SHA-256, and manually vetted for safety.  Table~\ref{tab:bankdata} and Table~\ref{tab:querytype} show summaries of our dataset. 
Of the nearly 78 million database operations captured, 61 million are not directly queries, but rather invocations of stored procedures.  The aim of this paper is to study query clustering specifically, so we ignore these queries.  
Of the the 17 million SQL statements in the trace, we were able to reliably parse 2.8 million.  Of these, we base our analysis on the 1.35 million syntactically valid \texttt{SELECT} queries.  

Our first observation is that of these 1.35 million queries, there are only 1614 different query skeletons.  The number of distinct structures that \sysname{} needs to cluster is quite small.  Furthermore, this number grows sub-linearly over time.  As shown in Table~\ref{tab:increasing_query_logs}, nearly half of all the skeletons arise in the first 1.5 hours of the trace.

\begin{table}[h]
  {\footnotesize
    \centering
    \begin{tabular}{|l|l|}
      \hline
      Not a query & 61,041,543\\
      \hline
      Un-parsable queries & 14,138,723\\
      \hline
      Parsable queries & 2,818,719\\
      \hline
      Total & 77,998,985\\
      \hline
    \end{tabular}
    \caption{Breakdown of bank data set.}
    \label{tab:bankdata}
  }
\end{table}

In the set of parsable queries in Table~\ref{tab:bankdata}, the distribution of different types of queries is shown in Table~\ref{tab:querytype}. 
\begin{table}[h]
  {\footnotesize
    \centering
    \begin{tabular}{|l|r|}
      \hline
      SELECT & 1,349,861\\
      \hline
      UNION & 1,306\\
      \hline
      INSERT & 1,173,140\\
      \hline
      UPDATE & 288,098\\
      \hline
      DELETE & 6,314\\
      \hline
      Total & 2,818,719\\
      \hline
    \end{tabular}
    \caption{Distribution of query types in bank dataset.}
    \label{tab:querytype}
  }
\end{table}

\subsection{Feasibility}

Our first goal is to understand whether clustering queries based on structural similarity, and in particular our use of the generalized W-L algorithm, produces meaningful results: Is log summarization even feasible in the first place?
In order to demonstrate the feasibility of log summarization, we compare \sysname{}'s query clustering with an ``intuitive'' query similarity metric applied.

To make a meaningful manual clustering tractable, we selected a sample of 140 query skeletons (about 10\%) and grouped them manually by attempting to infer the query's ``intent'' and creating clusters of queries with similar intents.

\begin{example}
\label{example:query_intent}
An example of two structurally similar queries in the set of $140$ query skeletons that are grouped together in manual clustering:
\begin{lstlisting}[language=SQL, showspaces=false, tabsize=1]
SELECT historytran.* 
FROM historytran LEFT JOIN feestate AS feestate 
	ON feestate.seqhistorytran = historytran.seq 
WHERE (historytran.caseid = '') AND 
	isnull(feestate.rechargestate, '') IN ('', '') 
ORDER BY historytran.txdate DESC, 
	historytran.txtime DESC
\end{lstlisting}
\begin{lstlisting}[language=SQL, showspaces=false, tabsize=1]
SELECT historytran.* 
FROM historytran LEFT JOIN feestate AS feestate 
	ON feestate.seqhistorytran = historytran.seq 
WHERE (historytran.caseid = '') AND 
	feestate.reversestate = '' AND 
	isnull(feestate.rechargestate, '') NOT IN ('', '') 
ORDER BY historytran.txdate DESC, 
	historytran.txtime DESC
\end{lstlisting}
When observing these queries, one can notice that they are exactly the same, except that second query contains \lstinline[language=SQL]{WHERE} expression with an additional \lstinline[language=SQL]{AND} phrase and $3^{rd}$ term is \lstinline[language=SQL]{NOT IN} instead of \lstinline[language=SQL]{IN} . From these two queries, a human inspector judged that they share similar intent and group them into one cluster with the reason that they select the same set of columns, show the results in same order, get data from same tables, and have similar \lstinline[language=SQL]{WHERE} condition.
\end{example}

The chance of false positives in a sample is governed by the birthday paradox; The chance of picking similar, but unrelated skeletons for our sample grows with the square of the sample size.  
To make our sampled results meaningful and increase the chance of false positives, we generated the 140 skeleton sample biased towards skeletons that share common terms and features (e.g., queries over the same relations).

\begin{figure}[h]
\centering
\includegraphics[width=\columnwidth]{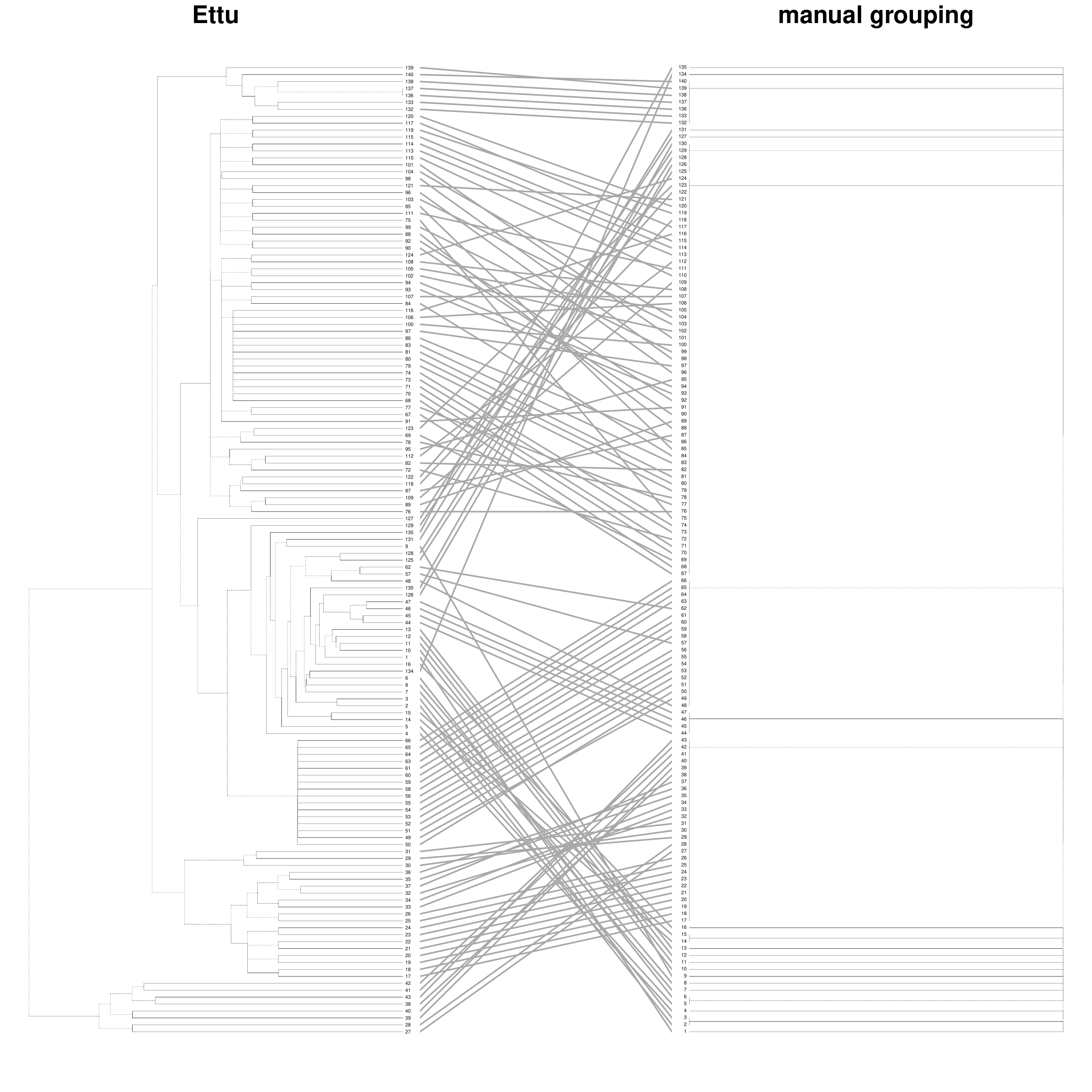}
\caption{Correspondences between hierarchical clustering obtained from \sysname{} and from manual grouping.}
\label{fig:tanglegram}
\end{figure}

We used \sysname{} to cluster the 140 query skeleton sample using hierarchical clustering and manually grouped the same set of queries.  By hand, we identified 23 
specific clusters of similar queries. 
To roughly visualize the correspondences between manual and automated clustering, Figure~\ref{fig:tanglegram} shows a tanglegram~\cite{galili2015dendextend} view of the clusters.
Ettu's clustering selections are shown as a dendrogram on the left of the tanglegram.  The x-coordinate of each branch signifies the distance between the clusters being merged.  The further to the right the branch appears, the more similar the clusters being merged.  Each leaf is one of the 140 query skeletons.
The manual clustering appears on the right of the tanglegram; Group assignments are fixed, and hence there are only two levels of the ``tree''.  Lines in the middle link skeletons in Ettu's clustering with the manual clustering.  

One measure of degree of correspondences between two trees is number of crossings of lines connecting two labels. The lower the number is, the higher correspondence between two trees is. However, aligning two trees so that the number of crossings is minimum is an NP-hard problem~\cite{buchin2008drawing}.  Figure~\ref{fig:tanglegram} was produced using the Dendextend R package~\cite{galili2015dendextend}, which minimizes the number of crossings but does not guarantee an optimal alignment. However, even with this figure, we qualitatively observe a large number of parallel lines (an indicator of good alignment between clustering results).

To measure the correspondence more quantitatively, we use a metric called entanglement~\cite{galili2015dendextend}, also implemented by Dendextend. Entanglement score ranges from 0 (no entanglement) and 1 (full entanglement). 
To compute an entanglement score, we first construct a vector for each clustering from the integer positions of each query in the dendrogram.  
The entanglement score is computed as the L-norm distance between these two vectors.
For the tanglegram in Figure~\ref{fig:tanglegram}, the L-2 entanglement score is 0.17, which indicates a high degree of correlation between the two clusterings.  
This demonstrates that \sysname{}'s distance metrics, based on structural similarity, do in fact correspond to more intuitive notions of query similarity.  

Surprisingly, a significant number of the alignment errors detected above are attributable (in retrospect) to human error rather than to \sysname{}.  After performing this experiment, we manually inspected the misalignments in Figure~\ref{fig:tanglegram} to better understand how \sysname{} was breaking.  In doing so, we found that many of the misalignments were actually mistakes in the manual grouping rather than the automated clustering algorithm. 
The reason for this error was that our manual clustering decisions were heavily biased by similarities between the first few words of each query: \texttt{SELECT} and the project targets.  
Conversely,  \sysname{}'s automated clustering is able to identify similarity in substructures as well as at the root of the query.

\subsection{Scalability}

Our second goal is to demonstrate the scalability of our proposed framework.  
There are two components to this framework: The initial clustering process which can be performed offline, and the interactive FP-tree explorer.  Concretely, we aim to show three properties of this framework.  First, the time taken by the offline component should be reasonable compared to the duration of the log.  The system is not useful if it takes the better part of a day to cluster a day's worth of queries.  Second, the added compute requirements of additional queries should likewise be reasonable.  Finally, the FP-tree explorer should be able to re-render a feature set at interactive speed.  

The bank query log starts at about 11 AM.  To test scalability, we vary the number of queries processed by truncating the log after varying lengths; After 1.5 hours, the log contains 250 thousand SELECT queries with 756 skeletons, just over 500 thousand queries after 3.3 hours, and so forth.  Full statistics for each truncated log are shown in Table \ref{tab:increasing_query_logs}.

Experiments were performed on a commodity laptop computer with a 2.2 GHz Intel Core i7 processor and 16 GB RAM memory running OS X 10.11.3. We used the R implementation of hierarchical clustering. The remaining components were implemented in Java using the single-threaded JDK 1.8.0. Reported running times are the average of 10 trials for each phase.

\begin{table}[h]
  {\footnotesize
    \centering
    \begin{tabular}{c|r|r|r}
    		 \textbf{End Time} & 
		 \textbf{Total Time} &
  		 \textbf{\texttt{SELECT}s} & 
		 \textbf{Skeletons} \\[0.5mm]
    		 \hline
		 &&&\\[-2mm]
    		 12:34 PM & 1.5 hours & 250,740 & 756\\
    		 14:24 PM & 3.3 hours & 512,981 & 946\\
    		 16:06 PM & 5 hours & 789,050 & 1,057\\
    		 18:27 PM & 7.3 hours & 972,511 & 1,102\\
    		 21:23 PM & 10.3 hours & 1,079,427 & 1,130\\
    		 00:43 AM & 13.6 hours & 1,176,582 & 1,256\\
    		 04:04 AM & 17 hours & 1,270,984 & 1,576\\
    		 06:28 AM & 19.4 hours & 1,349,861 & 1,614\\[0.5mm]
		 \hline
    \end{tabular}
    \caption{Summary of query log sizes by absolute time.}
    \label{tab:increasing_query_logs}
  }
\end{table}

\begin{figure}[h]
\centering
\includegraphics[width=\columnwidth]{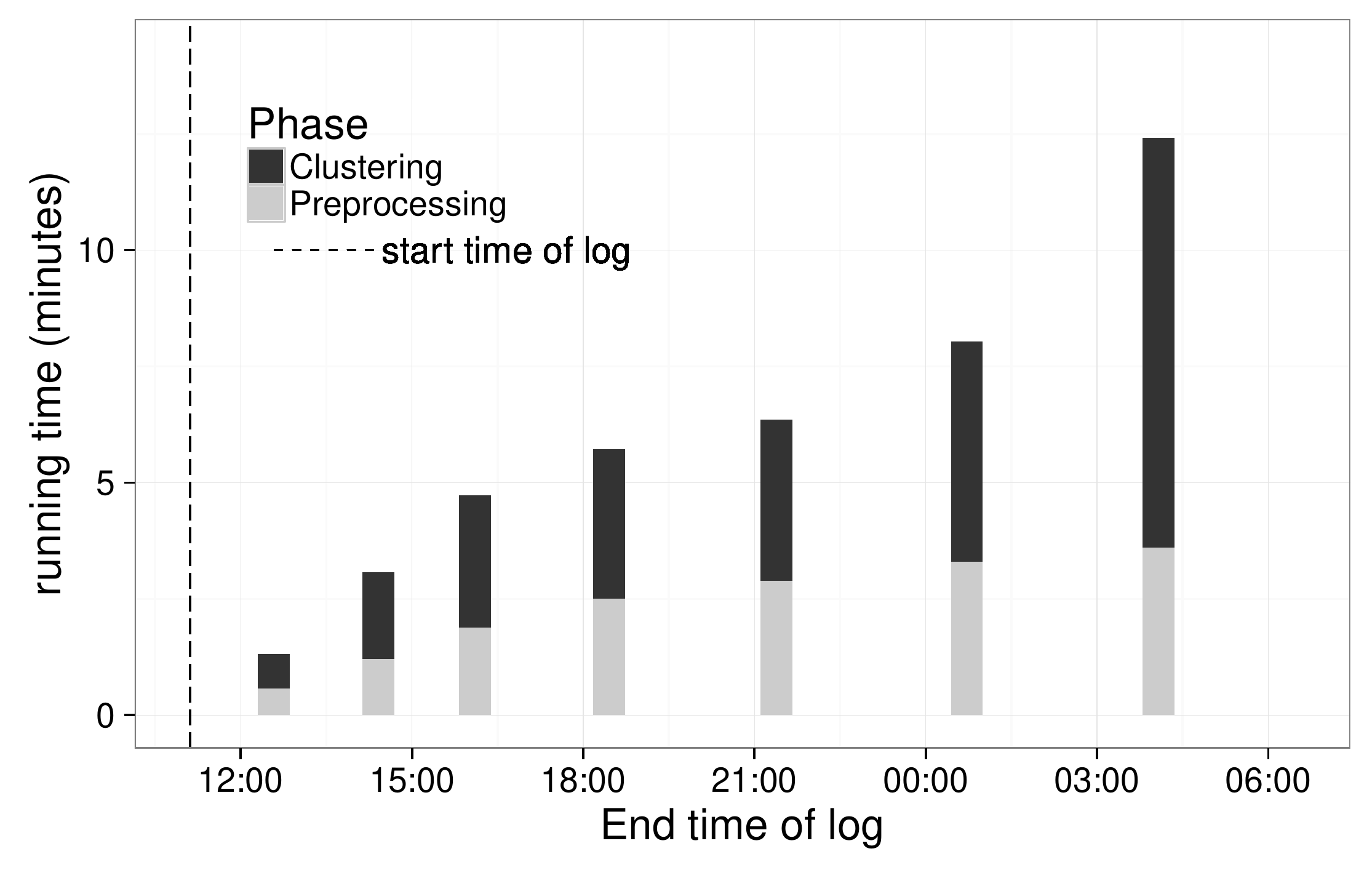}
\caption{Running time of preprocessing phase and clustering phase with respect to different log size.}
\label{fig:running_time}
\end{figure}

\begin{figure}[h]
\centering
\includegraphics[width=\columnwidth]{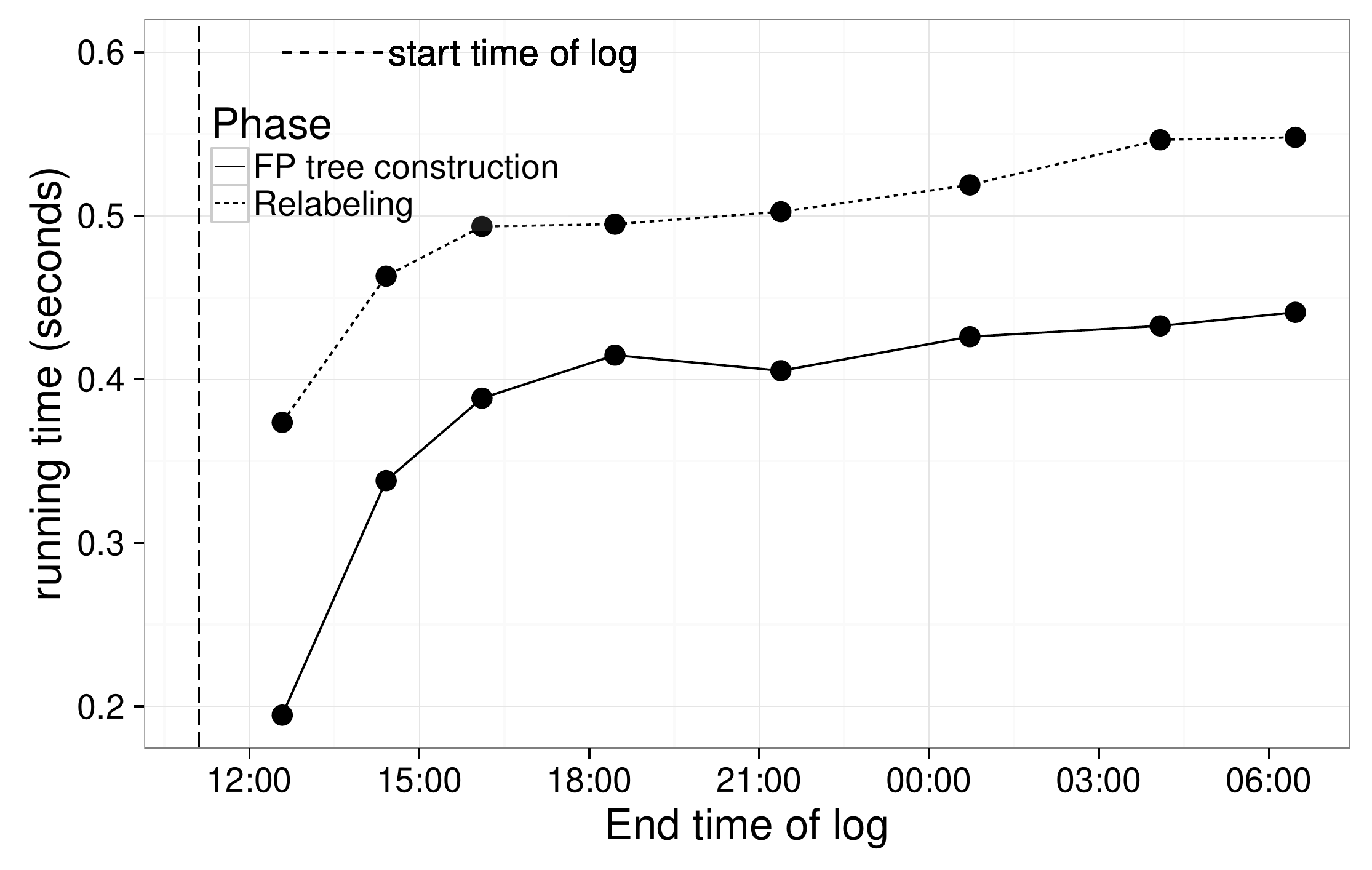}
\caption{Running time of relabeling phase and FP tree construction phase with respect to different log size.}
\label{fig:running_time_labeling}
\end{figure}

The clustering process is performed in three distinct phases: (1) A \textbf{preprocessing} phase where the full query log is rewritten into a bag of query skeletons, (2) A \textbf{relabeling} phase where the generalized W-L algorithm is applied to each query skeleton to extract feature vectors, (3) A \textbf{clustering} phase that actually groups queries by similarity, and finally (4) A \textbf{FP tree construction} phase in which each cluster of query skeletons is summarized using FP tree.  
Figure~\ref{fig:running_time} shows the running time for phases 1 and 3.  Phase 2 and phase 4 take almost no time by comparison and is shown independently in Figure~\ref{fig:running_time_labeling}.

The running time of phase 1 grows sub-linearly with the log size. We attribute this to the dominant costs being resource allocation associated with defining and indexing new query skeletons, and potentially JIT compilation overheads.  The running time of phase 2 has (relatively) high upfront costs, predominantly $\digest{\cdot}$ needing to index newly allocated feature identifiers, but quickly levels off into a linear scaling with the number of queries.  The running time of phase 3 grows super-linearly with the number of query skeletons. This is typical for clustering processes in general, which need to compute and evaluate a full ($O(N^2)$) pairwise distance matrix.  Fortunately, the number of skeletons grows sub-linearly with the number of queries, and the overall scaling by time is roughly linear; We expect that this growth curve would eventually become sub-linear for longer traces. Phase 4 runs quite fast because of relatively small number of query skeletons. 

With running time breaks down into steps as shown in Figure~\ref{fig:running_time} and Figure~\ref{fig:running_time_labeling}, we can observe that the bottleneck of our system is in hierarchical clustering in which running time grows quadratically fast as the log size grows.  However, our experiments, even with single-threaded on commodity hardware, were able to cluster the trace of parseable queries in under 15 minutes.  We can reasonably expect the clustering process to scale to even larger workloads, especially considering the embarrassingly parallelizable nature of clustering.


%% file: sections/4-relatedwork.tex
The basic idea behind \sysname{} is to profile normal user behavior, detect suspicious behavior using this information, and distinguish malicious behavior from benign intents~\cite{Kul2015Jowua}. 
Indeed, this idea is not new; there are many anomaly detection systems focusing on suspicious behavior of users.  
Specific examples include file access~\cite{wang2016fileaccess} and transfers~\cite{Arief2015Jowua}, online and social behavior~\cite{Gavai2015Jowua}, activities on a website~\cite{manavoglu2003network}, command-line statements~\cite{maxion2002masquerade} and SQL queries issued to a database~\cite{Mathew2010Raid}. 

\textbf{SQL queries as a resource:}
As the basic unit of interaction between a database and its users, the sequence of SQL queries that a user issues effectively models the user's behavior. 

One approach relies on the syntax of queries~\cite{Kamra2007SyntaxBased} and it permits fast query validation where they focus on detection of potential intrusions on database systems. In their paper, they introduce a mechanism which analyzes audit logs of databases with both defined user roles and undefined user roles. This system uses multiple techniques to attempt to detect the threats depending on the role distinction.
They use Naive Bayes Classifier and clustering techniques, k-centers and k-means, in their experiments to build user profiles.
The techniques were able to produce low false positives in experimental testing, but the false negative rates were high for both techniques. Their work shows that building user profiles from database logs has potential for detecting intrusions, especially in a system with defined roles.
\sysname{}, however, takes a different approach in creating the feature vectors exploiting subtree similarities instead of individual feature similarities, in order to address the problem when the syntax is much different but the main intent of the query is similar.

Another method is to use a data--centric approach, which performs better in detecting anomalies \cite{Mathew2010Raid}.
In the paper, a set of techniques are used to detect anomalies on the results of SQL queries rather than the queries themselves.
In doing this, the authors are able to improve both where the syntax--based methods miss a potentially harmful query and when there is a false positive.
These errors can occur in the syntax--based system when two queries with seemingly similar structures have vastly different results, as well as when two queries appear unrelated but actually return the exact same set of data.
The limitation of this approach is that the dataset should remain static for effective results but still the potential in the data--centric model is quite promising.
However, to evaluate the efficacy of such a system in a realistic way, access to data in a corporate database is required.

This brings us back to the problem that, many organizations do not allow third party softwares to access their sensitive data, let alone letting them to be used for research purposes~\cite{Kul2015Jowua}.
Hence, auditing SQL query logs of the databases and having a sense of what the query intends to do correctly gains a lot of importance.
Our eventual goal is extend our syntactic visualization strategy to support data--aware visualization and query auditing, in this work we focus on the core components of clustering queries and generating cluster visualizations.

\textbf{Query visualization:}
Generating query cluster visualizations is a complex task considering that even visualizing just one query accurately to help users understand the intent behind the query quickly is still a research challenge~\cite{gatterbauer2011databases}.

QueryViz~\cite{Danaparamita2011queryviz} addresses \textit{query interpretation} which is the problem of understanding the goal of the query by visualizing it. Query interpretation is usually as hard as writing a new query as the complexity of the query increases~\cite{gatterbauer2011databases}. It takes SQL query and the schema of the database; and parses the query, builds an AST and creates a graph for users to view it. The aim is to present queries as simple as possible for the users to understand them, as it is easier to understand the relationships and references in a query when it is graphically visualized.

Logos~\cite{Kokkalis:2012:LST:2213836.2213929}, on the other hand, is a system that has the ability to translate SQL queries into natural language equivalents. This technology works by creating graph representations of the query structure. These representations then use predefined relationships within the database schema to allow for construction of natural language expressions. Despite the high overhead of maintaining the relationships for the database, this technology shows promise in revealing the intent of user queries.

We take a similar approach as QueryViz~\cite{Danaparamita2011queryviz}, but we include the common features of all queries in the query representation, so that the users can see what features are unique to the query and what features are shared by all the queries in the query groups.

\textbf{Distinguishing queries:}
There are different approaches to understand the intents behind SQL queries and what a query returns.
Analyzing query intents can help in different research topics like creating indices, benchmark design, and masquerade detection in databases.

QueryScope~\cite{hu2008queryscope} starts with the hypothesis that visualization of queries would help identifying similar queries and distinguishing different ones from each other, hence it would help finding better tuning opportunities. The system provides a user interface to visualize SQL queries. This interface presents these queries as graphs in order to make the structure more understandable. To construct these graphs, the structure of the query is broken down into XML format prior to visualization. The proposed use of this interface is for tuning of database systems, although the breakdown of these queries proposes the potential for detecting similarity between different queries. Our approach is similar to the individual queries but we focus on describing query sets. While they directly focus on similarities of queries by finding the critical elements of the query using the table relationships and table cardinalities, we make use of the subtree similarities.

XData~\cite{chandra2015Data} is a system which uses a technique that allows SQL queries to be tested for mutations to the intended query. In doing this, queries that have slight differences and that look to return the same result as a correct query will be marked as incorrect if they produce different results. This technology is intended for use in grading student work, although other uses are possible.
In order to determine the correctness of a query, the correct query must have constraints created to represent possible mutations. The authors explain each aspect of a SQL query for which a mutation can occur, and for each they use tuples to store the possible mutations. By doing this, all possibly mutation combinations can be presented for comparison against questionable queries. In testing against student work, this technique was able to correctly identify over 95 percent of incorrect mutants presented.

A method for maintaining privacy in databases via auditing is presented in~\cite{Agrawal:2004:ACH:1316689.1316735} which discusses an auditing framework that allows for analysts to ensure compliance of certain privacy guarantees. Audit expressions are used in their paper to establish this notion of privacy in the database along with the concept of an indispensable tuple. This means that the removal of this tuple will affect the result of the query in question. By using this concept, the authors detail a methodology in which audit logs can be constructed that will reveal any privacy violations.


If the data for the target database is available, our proposed framework can be augmented by using methods proposed in above discussed data--centric approaches~\cite{chandra2015Data, Agrawal:2004:ACH:1316689.1316735}.



%% file: sections/5-futurework.tex
The focus of this paper is to summarize large query logs by clustering them according to their similarity with the hypothesis that the similarity in query structure corresponds to the intent of the query.
 We utilize  a revised Weisfeiler--Lehman algorithm to create features out of query logs, form feature vectors out of them, and cluster the similar queries together by their feature vectors with hierarchical clustering method.
Finally, we exploit FP Trees to create summaries of the clustered query sets. In our experiments, we show that
(1) the structural similarity of queries corresponds to the intent of the query by comparing the groupings performed by a human expert and our system, and
(2) even with a commodity laptop and single-threaded implementation, the process could be performed in under 15 minutes for a set of 1.35 million queries.


We plan to extend our work in several directions:
First, we will explore new feature weighting strategies and new labeling rules in order to capture the intent behind logged queries better.
Second, we will examine user interfaces that better present clusters of queries --- Different feature sorting strategies in an FP Tree in order to help the user distinguish important and irrelevant features, for example.

\begin{scenario}
Jane inspects the query group summaries presented by \sysname{} and labels these groups as safe, unsafe, and unknown. \sysname{} repeats the auditing process to elaborate the groups on unknown clusters, so that they can be labeled as safe and unsafe, too.
\end{scenario}

Third, further exploration on various kinds of statistics captured in FP Tree will help us in
determining the quality of the cluster, weighting of features, and identifying suspicious paths.
Lastly, we will investigate the effect of temporality on query clustering. 

